\documentclass[twocolumn,twocolappendix]{aastex631}

\usepackage{graphicx}
\usepackage{amssymb}
\usepackage[fleqn]{amsmath}
\usepackage{times}
\usepackage{subfigure}
\usepackage[T1]{fontenc}
\usepackage{ae,aecompl}
\usepackage{textcomp}

\renewcommand{\vec}[1]{\boldsymbol{#1}}

\begin{document}

\title{Inferring Stellar Densities with Flexible Models I: The Distribution of RR Lyrae in the Milky Way with \textit{Gaia} DR3}

\author[0000-0001-7297-8508]{Madeline Lucey}
\affiliation{
Department of Physics \& Astronomy, University of Pennsylvania, 209 S 33rd St., Philadelphia, PA 19104, USA }

\author[0000-0002-6330-2394]{Cecilia Mateu}
\affiliation{Departamento de Astronomía, Instituto de Física, Universidad de la República Iguá 4225, CP 11400 Montevideo, Uruguay}

\author[0000-0003-0872-7098]{Adrian M. Price-Whelan}
\affiliation{Center for Computational Astrophysics, Flatiron Institute, 162 5th Avenue, New York, NY 10010, USA}

\author[0000-0003-2866-9403]{David W. Hogg}
\affiliation{Center for Computational Astrophysics, Flatiron Institute, 162 5th Avenue, New York, NY 10010, USA}
\affiliation{Max Planck Institute for Astronomy, Königstuhl 17, D-69117 Heidelberg, Germany}
\affiliation{Center for Cosmology and Particle Physics, Department of Physics, New York University, 726 Broadway, New York, NY 10003, USA}

\author[0000-0003-4996-9069]{Hans-Walter Rix}
\affiliation{Max Planck Institute for Astronomy, Königstuhl 17, D-69117 Heidelberg, Germany}

\author[0000-0003-3939-3297]{Robyn E. Sanderson}
\affiliation{Department of Physics \& Astronomy, University of Pennsylvania, 209 S 33rd St., Philadelphia, PA 19104, USA}

\begin{abstract}

Understanding the formation and evolutionary history of the Milky Way requires detailed mapping of its stellar components, which preserve fossil records of the Galaxy's assembly through cosmic time. RR Lyrae stars are particularly well-suited for this endeavor, as they are old, standard candle variables that probe the Galaxy's earliest formation epochs. In this work, we employ a hierarchical Bayesian Gaussian Mixture Model (GMM) to characterize the three-dimensional density distribution of RR Lyrae stars in the Milky Way over the galactocentric radius ($R$) of $\approx 0.2-120~\rm{kpc}$. This approach provides a flexible framework for modeling complex stellar distributions, particularly in the inner Galaxy  where the bulge, disk, and halo components overlap. Our analysis reveals that the inner Galaxy ($R\lesssim10~\rm{kpc}$) is dominated by a distinct prolate stellar population with axis ratio $q$=1.31. Consistent with previous work, we find the halo follows a $r^{-4}$ power-law profile that flattens within 12 kpc of the Galactic center. We also confirm the outer halo ($R\gtrsim10~\rm{kpc}$) is oblate $q$=0.70 with a tilt angle of $18^{\circ}$. We report for the first time that this tilt aligns the halo major axis in the direction of the Sagittarius dwarf galaxy. These results establish GMMs as an effective and flexible tool for modeling Galactic structure and provide new constraints on the distribution of old stars in the inner Galaxy. 

\end{abstract}



\section{Introduction} \label{sec:intro}

With the billions of resolvable stars in the Milky Way, we can study the formation and evolutionary history of our Galaxy in exquisite detail. The Galaxy's past structure is chronicled in the distribution of its resolved stellar populations \citep{Grebel1999,Feast2000,Freeman2002,Sandage2006,Catchpole2016}. By studying the structure of a stellar population with a constrained age distribution we obtain a glimpse into the Galaxy's dynamical history since that population's birth.  As different galaxy formation processes lead to different correlations between stellar age and dynamics,  we can use these observed patterns to distinguish between formation scenarios \citep{Eggen1962,Searle1978,Freeman1987,Molla1996,Walker1996,Bland-Hawthorn2016,Helmi2020,Deason2024}. 

Among the most valuable tracers for this purpose, RR Lyrae (RRL) are pulsating horizontal branch stars that are predominantly old ($\geq$10 Gyr) and metal-poor \citep[$\rm{[Fe/H]\lesssim-0.5}$;][]{Smith1995}, making them an effective probe for the Galaxy's earliest evolution. Furthermore, they are an ideal population for studying Galactic structure because their standard candle nature gives precise distance estimates as low as 1-2\% when both infrared and optical photometry is available \citep{Madore2012,Neeley2017,Sesar2017a,Mullen2023}. Even in high extinction regions, RRL are still excellent distance indicators because they are color standards; they have the same effective temperature and therefore intrinsic color during their minimum light phase \citep{Sturch1966,Guldenschuh2005,Kunder2010,Vivas2017}.

These exceptional properties have made RRL valuable for studying many aspects of the Galaxy's structure, including the Galactic disk. The disk is typically modeled with two components: the older thick disk and the younger thin disk \citep{Gilmore1983,Wyse1995,Bensby2011,Fuhrmann2008,Reddy2006}. Given that they are generally old, RRL have been primarily used to study the Galactic thick disk \citep{Layden1995,Amrose2001,Kinemuchi2006,Mateu2012}. Overall, the thick disk red giant branch (RGB) and red clump (RC) density distribution is consistent with results for RRL with most estimates converging around an exponential scale height ($h_z$) of $\approx600-800$ pc and scale length ($h_R$) of $\approx 2.1$ kpc \citep{Bovy2012a,Bovy2012b,Bovy2016b,Robin2014,Mateu2018,Tkachenko2025}.

The relationship between RRL and the thin disk, however, has proven more complex. Although, metal-rich RRL stars with thin disk kinematics were found by \citet{Layden1995} and as early as \citet{Strugnell1986}, other studies   were unable to confirm the association of these stars to the thin disk and suggested they might belong to the metal-rich tail of the thick disk \citep{Martin1998}.
Recent work has revealed a significant fraction of disk RRL with thin disk kinematics \citep{Iorio2021},  sparking a debate about their origins \citep{Chadid2017,Marsakov2018,Marsakov2019,Zinn2020,Prudil2020}. One hypothesis is that these stars have an accreted origin \citep{Feuillet2022}. Another possibility is that the thin disk RRL are actually an intermediate-age population and that it could be the result of binary evolution \citep{Iorio2021,Bobrick2024,Cuevas-Otahola2024,Cabrera-Gadea2024,Zhang2025}. 

Beyond the Galactic disk, the metal-poor RRL serve as ideal tracers for the stellar density distribution of the Galactic halo, which is thought to consist of ancient accreted galaxies \citep{De_lucia2008,Deason2013}. Generally, results for the halo density distribution of RRL are consistent with the total stellar distribution which are both most commonly  modeled as a broken power law with a slope between $\alpha\approx-1.5$ to $-3$ for the inner halo and a slope of $\alpha\approx-4$ for the outer halo with the break radius around $\approx20-30$ kpc \citep{Sesar2007,Watkins2009,Deason2011,Sesar2013,Faccioli2014,Zinn2014,Das2016,Hernitschek2018,Iorio2018,Chen2023,Lane2023}. However, other works have found doubly broken power-law slopes with the second break radius of $\approx 10~\rm{kpc}$ \citep{Han2022,Yang2022}. Recent works have also found the slope may be distinct along different lines-of-sight in the Galaxy \citep{Hernitschek2018,Amarante2024,Medina2024}.  Works focusing on the very distant outer halo ($R\gtrsim100~\rm{kpc}$) have found similar slopes of $\alpha\approx-4$ \citep{Thomas2018,Medina2018,Stringer2021,Feng2024}, although some have found a break radius at $R\approx160-210~\rm{kpc}$ \citep{Fukushima2018,Fukushima2019}. None of these previous works include data within 5~kpc of the Galactic center with only a few of them including data with Galactocentric radii <10~kpc. There is also an interesting trend in the literature where works that include data at smaller radii tend to measure flatter slopes for the inner halo. Furthermore, previous works find that the halo is flattened with $q\approx$0.6-0.7 where $R=\sqrt{X^2+Y^2+(Z/q)^2}$ \citep{Deason2011,Iorio2018,Han2022} with most recent work also measuring a tilt in the halo with respect to the Galactic disk plane \citep{Iorio2019,Han2022}. There is also work which finds the flattening varies with Galactocentric radius \citep{Xue2015,Xu2018,Iorio2018,Hernitschek2018}.

Perhaps the most challenging region to characterize, the inner Galaxy contains many overlapping components (halo, disk and bulge) and complex structures (the Galactic bar and X-shape). Whether the bulge is a distinct component from the inner halo is currently debated \citep{Lucey2021,Ardern-Arentsen2024}. \citet{Pietrukowicz2015} found the RRL density distribution in the inner Galaxy (0.2-3 kpc) is consistent with the measurements of the inner halo with a power law slope of $\alpha\approx$-3 \citep{PerezVillegas2017}. There is also evidence that a population of RRL in the inner Galaxy trace the bar structure \citep{Kunder2020}.

In this work, we simultaneously model the RRL density distribution of the Milky Way bulge, halo and disk. In order to flexibly model all these structures, we utilize a Gaussian Mixture Model 
(GMM). GMMs have previously been used to model stellar density distributions of external galaxies, including to approximate power laws \citep{Cappellari2002,Hogg2013,Miller2021}. However, this work is the first time this method has been applied to measure the  stellar density distribution of the Milky Way. 

This publication is laid out as follows. In Section \ref{sec:Data} we describe the sample of RRL, including parameterizing the selection function and calculating the distance modulus. We introduce our hierarchical Bayesian model of Gaussian mixtures in Section \ref{sec:method}. We present our resulting density distribution of RRL and a comparison to previous work in Section \ref{sec:Res}. Last, in Section \ref{sec:Sum}, we summarize this work and the conclusions.

\section{RR Lyrae Sample}\label{sec:Data}

\begin{figure*}
    \centering
    \includegraphics[width=\linewidth]{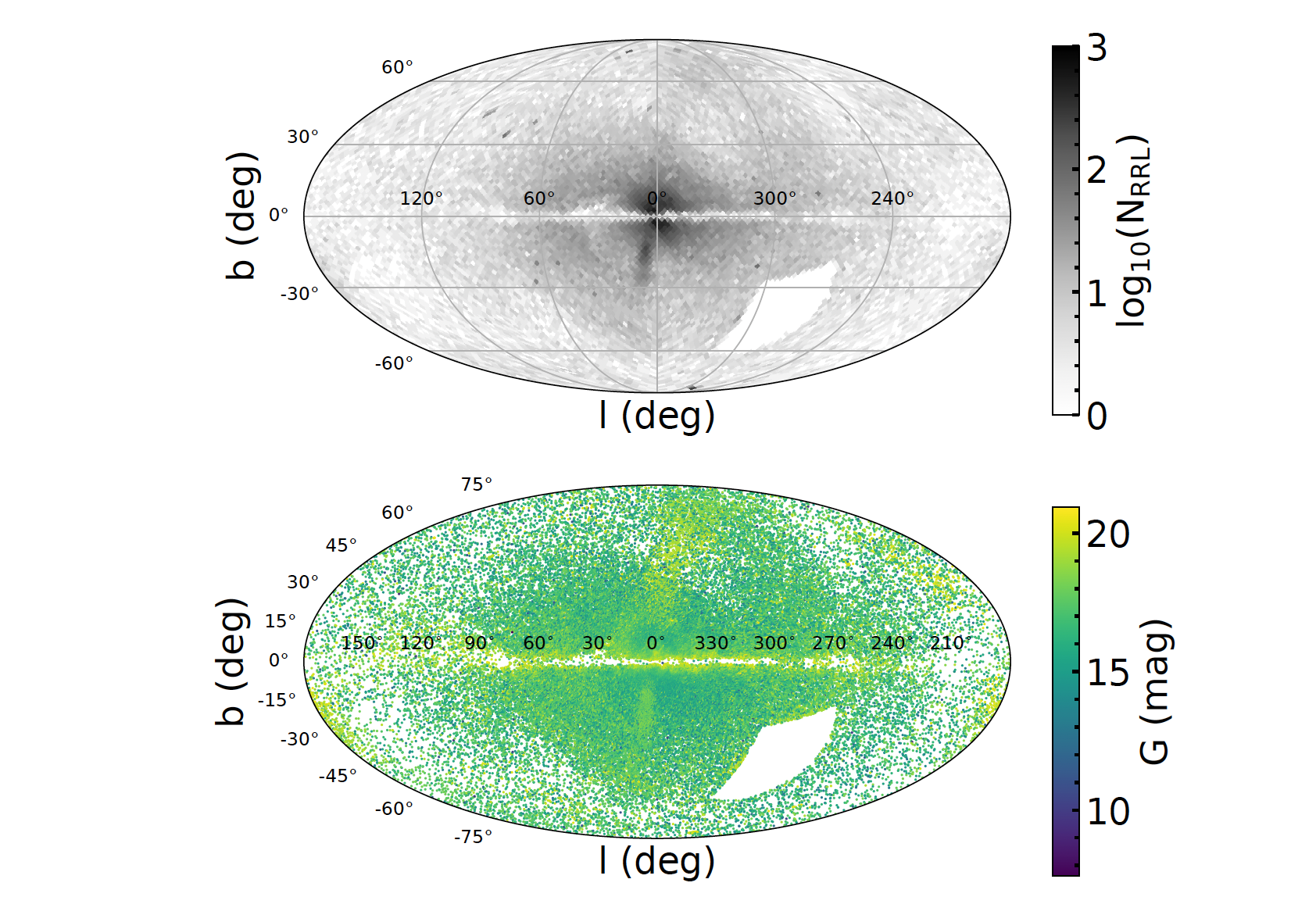}
    \caption{The sky distribution in Galactic coordinates ($l,b$) of our RRL sample. The top panel shows the log density for the 105,325 RRL we use in this work, after removing the LMC and SMC with a cut in RA and DEC. The bottom panel shows the sky distribution of our final sample colored by the \textit{Gaia} G magnitude, which, because of their standard-candle nature, is strongly correlated with the distance. }
    \label{fig:sky}
\end{figure*}

Large samples of RRL stars are effective tools for studying Galactic structure given that their standard candle nature provides precise distance estimates. In this work, we use a catalog compiled from the three largest public surveys of RRL stars: \textit{Gaia} DR3 Specific Objects Study \citep[SOS,][]{Clementini2023_SOS_GaiaDR3} ASAS-SN \citep{Jayasinghe2019} and PanSTARRS1 \citep[PS1,][]{Sesar2017c}. The quality cuts applied to each catalog are described in detail in Sec.~2 of \citet{Mateu2020}, as well as the cross-matching strategy used to match the known RRL from each catalog to a source in the \textit{Gaia} DR3 \verb|gaia_source| table. Duplicate sources were removed based on \textit{Gaia}’s \verb|source_id|, resulting in a base catalog with a total of 288,928 RRL.

Distances to the RRL stars were computed using the Period-Wesenheit-Metallicity (\emph{PWZ}) relation  from \citet{Neeley2019} for the \textit{Gaia} reddening-free Wesenheit index $W_{G,BP-RP}$, and photometric metallicities from \citet{Li2023}. These photometric metallicities were computed from the G-band light curve $\phi_{31}$ Fourier coefficient and re-calibrated by \citet{Li2023} to correct for systematic effects present in the photometric metallicities reported in the SOS catalog.  
Finally, the catalog was restricted to RRL stars with $\mathrm{RUWE}<1.4$ and $\verb|phot_bp_rp_excess_factor|<3$
to ensure, respectively, the quality of the astrometric parameters and of the BP/RP photometry used in the distance calculation; and filtering the catalog to stars with photometric metallicities available results in a final catalog of 128,353 RRL stars, comprised of 108,295, 19,986 and 72 RRL of types \emph{ab} (fundamental mode), \emph{c} (first overtone) and \emph{d} (double-mode). These stars have a heliocentric distance range of $0.2-140$ kpc, corresponding to a galactocentric distance range of $0.2-120$ kpc. 

We present the sky distribution of our sample in Galactic coordinates ($l,b$) in Figure \ref{fig:sky}.  To remove contamination from the two most massive dwarf galaxies orbiting the Milky Way, the LMC and SMC, we remove stars with $\mathrm{RA}<110^{\circ}$ and $\mathrm{Dec.}<-55^{\circ}$. This is the simplest way to remove these major structures from the sample without complicated selection function effects from distance uncertainties. The sky distribution of the resulting 102,761 RRL is shown in the top panel of Figure \ref{fig:sky}.

The bottom panel of Figure \ref{fig:sky} shows the sky distribution of our final sample colored by the observed G magnitude. The Sagittarius dwarf galaxy stream is visible as a streak of stars with high G magnitudes passing vertically through the Galactic center at $(l,b) = (0,0)^{\circ}.$ However, as this is much lower density than the LMC and SMC we do not find it necessary to remove from the sample. Along the disk plane at $b=0^{\circ}$, high levels of dust extinction cause higher G magnitudes as well as gaps in the RRL distribution. We account for this effect in our selection function which is described in the following Section.

\subsection{Selection Function} \label{sec:SelFunc}

\begin{figure*}
    \centering
    \includegraphics[width=\linewidth]{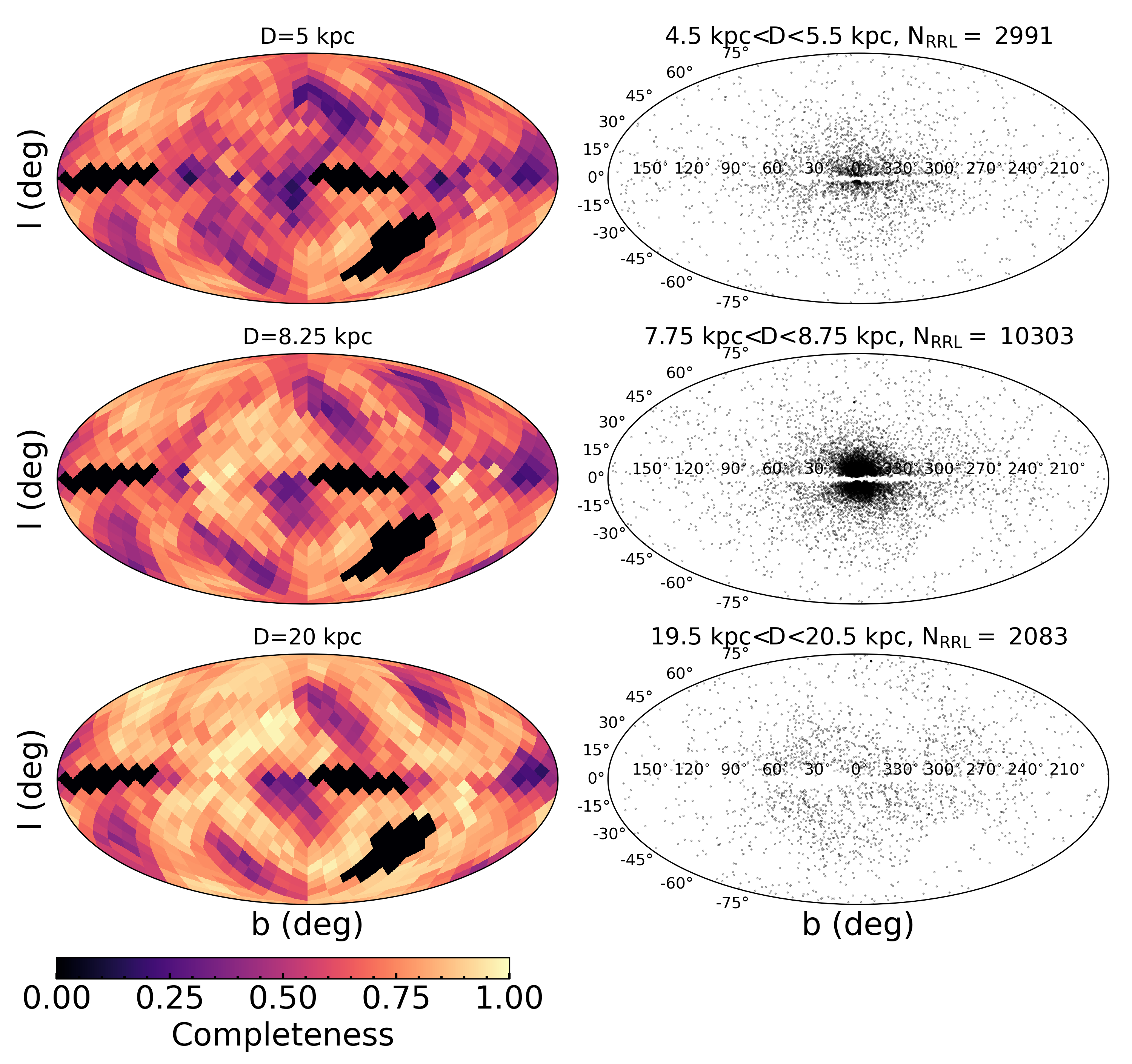}
    \caption{Empirical Completeness Maps of the RRL sample in Galactic Coordinates at heliocentric distances of 5 kpc (top left), 10 kpc (middle left), and 20 kpc (bottom left). We show corresponding slices of the data in heliocentric distance bins of 1 kpc in the right panels. At small distances, the completeness is degraded because of the bright limits of surveys. The dominant structure in the completeness map is consistent with the \textit{Gaia} scanning law.  }
    \label{fig:sf}
\end{figure*}

In order to account for the observational biases of our RRL sample, we need to quantify the selection function by estimating the completeness of our sample as a function of sky position and heliocentric distance. We estimate the completeness following the procedure described in detail in \citet{Mateu2020} and introduced by \citet{Rybizki2018}. The method uses two independent catalogs, in our case \textit{Gaia} and ASAS+PS1, to empirically determine the completeness for each catalog based on the number of stars in both catalogs as well as in each. Completeness is estimated separately for RRL of types \emph{ab} and \emph{c}\footnote{For these purposes we consider the very few RR\emph{d} as RR\emph{ab} since they are too few for a completeness map to be estimated.}, since RR\emph{c} samples are well known to be more affected by incompleteness as these stars have lower photometric amplitudes and are more prone to misclassification \citep[see e.g.][]{Mateu2020,Sesar2017c}.   In \citet{Mateu2020} completeness maps were presented for the \textit{Gaia} DR2 SOS, ASAS-SN and PS1 RRL catalogs, and an update presented in \citet{Mateu2024} for \textit{Gaia} DR3 and the combined \textit{Gaia}+ASAS+PS1. In the latter, completeness maps using only the RUWE and \verb|phot_bp_rp_excess_factor|   (BEP) filters were reported. Here we compute the completeness (or selection function) maps including also the more stringent availability filter on the \citet{Li2023} photometric metallicity (or equivalently \textit{Gaia}~DR3's). The resulting selection function maps with RUWE, BEP and photometric metallicities filters are publicly available in \citet{mateu_rrl_completeness_2025}. The tools to generate this and other similar selection functions are also available at the \verb|rrl_completeness| GitHub repository\footnote{\url{https://github.com/cmateu/rrl_completeness}} . 

In the left panels of Figure~\ref{fig:sf} we present images of the resulting selection function in Galactic coordinates evaluated at three heliocentric distances (5, 10, and 20 kpc from top to bottom). The right panels show corresponding 1 kpc slices of the data. The dominant structure is a result of the \textit{Gaia} scanning law \citep[see Figure 10 in][]{Cantat-Gaudin2023} where the completeness is lowest in regions with the fewest scans. At small heliocentric distances, the completeness is lower due to the bright limits of the surveys, while the faint limits impact the completeness at large heliocentric distances.

\section{Hierarchical Bayesian Model of Multi-Gaussian Galaxy } \label{sec:method}

The aim of this work is to find the maximum likelihood estimate for a given RRL density model described by parameters, $\vec{\theta}$, given our RRL distance catalog and selection function ($S(l,b,d)$). 
Specifically, we use a hierarchical Bayesian model to infer the true underlying stellar density distribution $n(\vec{x} \,;\, \vec{\theta})$, where $\vec{\theta}$ are the parameters of our density model.
We transform true 3D Galactocentric positions $\vec{x}$ to observed distances ($d$) and sky positions ($l,b$) and marginalize over the distance uncertainty as follows:
\begin{equation} \label{eqn:1}
\begin{split}
 \mathcal{L}(\vec{\theta}) 
& \propto \prod_{i=1}^N 
    P(l_{i},b_{i},d_{i}|\vec{\theta}, \vec{x}_i) \, 
    S(l_i,b_i,d_i) \hfill  \\
  &  \propto \prod_{i=1}^N  
    \int_{d_{min}}^{d_{max}}
        d\vec{x}_i \,
        P(l_{i},b_{i},d_{i}|\vec{x_i}) \,
        P(\vec{x_i}|\vec{\theta}) \,
        S(l_i,b_i,d_i)
 \end{split}   
\end{equation}
where 
\begin{equation}
    P(\vec{x}_i | \vec{\theta}) = \frac{1}{N} n(\vec{x}_i \,;\, \vec{\theta})
\end{equation}
is a normalized version of our density function, and the index $i$ tracks each of the $N$ RRL stars we use as data.

To perform the transformation to and from Galactocentric positions, we assume a right-handed coordinate system with the Sun's position at \citep[$-8.25$~kpc, 0, 20.8~pc;][]{Gravity2021,Leung2023}. The measured sky positions ($l,b$) of are significantly more precise than the distances ($d$) so the integral over 3D position simply becomes a 1D integral over the distance. For computational efficiency, this integral is calculated via a Riemann sum from $d_{min}$ to $d_{max}$, which are defined as follows: 
\begin{align}
d_{min} &= \min \begin{cases}
    d-3\sigma_{d}~[\rm{kpc}]\\
    10^{-3}~\rm{kpc}
\end{cases} \\
d_{max} &=  d+3\sigma_{d}~[\rm{kpc}]
\end{align}
where $\sigma_d$ is the uncertainty on the distance. The choice of 3$\sigma_d$ in the integration bounds is arbitrary, but sufficient given the Gaussian distance uncertainties, as this captures 99.7\% of the probability density. To prevent negative distance values in the case of $3\sigma_d>d$, a minimum distance of $10^{-3}$ kpc is chosen as an adequately small bound given that it is two orders of magnitude less than the smallest distance in our sample (0.28 kpc).

The $P(\vec{x}_i|\vec{\theta})$ term in Equation \ref{eqn:1} is our density model of the Galaxy. In this work, we use a GMM to describe the Galactic stellar density because it is a normalized probability density distribution unlike many of the power law models typically used to describe stellar halos \citep[e.g.,][]{Deason2011,Iorio2018,Han2022}. Furthermore, GMMs can accurately approximate many power law functions with only a few components \citep{Cappellari2002,Hogg2013,Miller2021}. While the choice of GMM is not physically-motivated, it offers numerous mathematical advantages \citep{Bendinelli1991,Bovy2011}, including an analytical solution to the Jean's equations \citep{Cappellari2008}, and better fits to real data where simple parametric models fail to recover the full complexity of galactic structure \citep{Emsellem1994}.

We fit two GMMs which are then combined with weights that sum to 1 as our complete density model. All Gaussian components of our model are centered on (0,0,0) kpc corresponding to the Galactic center. One of the GMMs is set to model the inner Galaxy, while the other models the halo. For both, we set the covariance matrices so that they increase by a factor of two between components. The difference between the inner Galaxy and halo model lies in the scaling factor ($k$) applied to the covariance matrices. For the inner Galaxy GMM $k=0.1$~kpc, while $k=1.5$~kpc for the halo GMM. The inner Galaxy GMM has 5 components while the halo GMM has 6. Using 6--11 total GMM components with logarithmically scaled widths is standard for modeling galaxy profiles \citep{Cappellari2002,Hogg2013,Miller2021}. Our method adds additional flexibility in allowing the halo to be flattened in a direction different than in the inner Galaxy. To allow flexibility in the model for exactly where the halo GMM begins to dominate the density profile, we ensure overlap in that the smallest halo GMM component has $\sigma_{xy}$=1.5~kpc and the largest inner Galaxy component has $\sigma_{xy}=$1.6~kpc.  Each GMM also includes its own flattening parameter ($q$) applied to the z-coordinate, which is nominally perpendicular to the Galactic disk. Therefore, for both GMMs, the non-rotated covariance matrix for the $jth$ component is as follows:

\begin{equation} \label{eq:halo}
cov_j =  \begin{bmatrix}
\sigma_{xy}^2 & 0 & 0 \\
0 & \sigma_{xy}^2 & 0 \\ 
0 & 0 & \sigma_{z}^2
\end{bmatrix} =  \begin{bmatrix}
k2^j & 0 & 0 \\
0 & k2^j & 0 \\ 
0 & 0 & kq2^j
\end{bmatrix}
\end{equation}

However, there is increasing evidence the the Galactic halo of the Milky Way may be rotated with respect to the Galactic disk \citep{Iorio2019,Han2022,Nibauer2025}. In other words, the Galactic halo may not be flattened in a direction directly perpendicular to the Galactic disk ($z$) but instead along another direction. To investigate this, we develop a rotated halo model. Specifically, we transform the halo components' covariance matrices with the following rotation matrix: 

\begin{equation} \label{eq:rot}
    \begin{split}
        R_{\rm rot} = \begin{bmatrix}
            \cos{\phi}\cos{\alpha}&-\sin{\phi}&\cos{\phi}\sin{\alpha} \\
            \sin{\phi}\cos{\alpha}&\cos{\phi}&\sin{\phi}\sin{\alpha} \\
            -\sin{\alpha}&0&\cos{\alpha}
        \end{bmatrix}  \\
    \end{split}
\end{equation}

where the new z-axis now points in the direction of the vector [$\sin{\alpha}\cos{\phi},\sin{\alpha}\sin{\phi},\cos{\alpha}$] in the non-rotated coordinate frame.  

In total, we use the following function form for the GMM of the density of RRL: 
\begin{equation} \label{eqn:model}
\begin{split}
    n(\vec{x}_i|\vec{\theta})&= \\  & (1-A)
        \sum_{j=1}^{6} A_{H,j} \, \mathcal{N}(\vec{x}_i|\vec{0},R_{\rm rot}cov_j(k=2.0~\rm{kpc})R_{\rm rot}^T) \\
    & + A \sum_{j=1}^5 A_{IG,j} \, \mathcal{N}(\vec{x}_i|\vec{0},cov_j(k=0.1~\rm{kpc}))
\end{split}
\end{equation}
where $A$ is the amplitude for the inner Galaxy GMM, $A_{H,j}$ is the amplitude for each halo component and $A_{IG,j}$ is the amplitude for each inner Galaxy component. 

We utilize the stick breaking transform to fit the amplitudes of the GMM components. For complete details of the transform we refer the reader to \citet{{StanReferenceManual}}. In short, it transforms $J$ GMM amplitudes to $J-1$ unconstrained parameters. This leads to better convergence because it reduces the dimensionality and ensures the amplitudes sum to 1. With this transform, our model has a total of 14 free parameters (10 amplitudes, 2 flattening parameters ($q$), and 2 angles) to fit.

\subsection{Inference} \label{sec:inf}

As our hierarchical Bayesian model is computationally expensive with high dimensionality (14 free parameters), we require fast, scalable optimizers and posterior estimators.  For fast maximum likelihood estimates, we use an the adafactor optimizer to find the minimum of the negative log-likelihood . We choose adafactor because it has similar benefits to the Adam optimizer (efficient stochastic gradient descent) with lower memory cost \citep{Shazeer2018} which is crucial in this work given that all of the data has to be held in memory at once in order to calculate the full likelihood. To estimate the uncertainties, we calculate the covariance matrix at the maximum likelihood estimate (MLE) using the negative inverse Hessian.

\section{The Galactic density distribution of RR Lyrae}\label{sec:Res}
\begin{figure*}
    \centering
    \includegraphics[width=\linewidth]{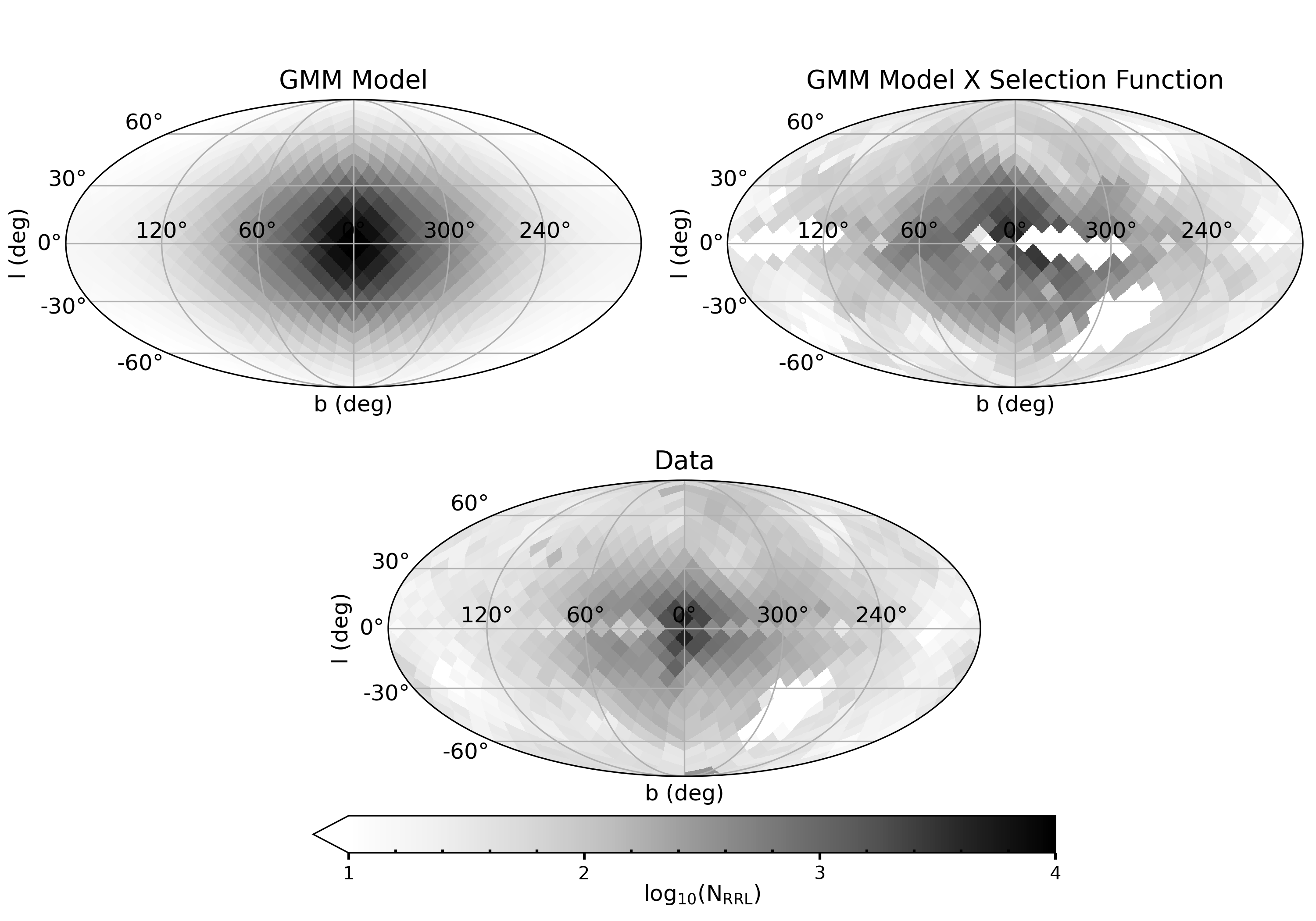}
    \caption{ Comparison of the sky projected densities of our model to the data in Galactic coordinates ($l, b$). The top left panel is sky projection of our model where the darkest patches indicate the highest density of RRL. The top right panel shows the same but convolved with our model for the selection function. The bottom panel shows the equivalent sky projection of the RRL sample used in this work. In general, the sky projected model is similar to the data when convolved with the selection.   }
    \label{fig:skymodel}
\end{figure*}

\begin{table}
\caption{Inferred Parameters of GMM Model}
\label{tab:table1}
\begin{tabular}{ccc}
\hline\hline
 & Inner Galaxy & Halo \\
 \hline
  q  & $1.311\pm 0.002$ &$0.697\pm 0.002$ \\ 
  
  $\rm{\alpha}~(^{\circ})$ & N/A &  $17.7\pm0.7$ \\
  $\rm{\phi}~(^{\circ})$   & N/A &  $-27.2 \pm 2.4$\\ 
\hline
Amplitudes for GMM & &  \\
\hline \hline
A &0.37 & \\

$j=$ & $A_{IG,j}$ & $A_{H,j}$\\
\hline
1 & $7.00 \times 10^{-2}$ & $9.99\times 10^{-5}$  \\ 
2 & $6.53\times 10^{-2} $&$1.14\times 10^{-4}$ \\ 
3 & $6.03\times 10^{-2}$ & $5.60\times 10^{-2}$   \\ 
4 & $5.05\times 10^{-2}$ & $7.70\times 10^{-1}$\\ 
5 &$ 7.53 \times 10^{-1}$ & $1.32\times 10^{-1}$ \\ 
6 & & $4.25 \times 10^{-2}$  \\ 
\hline
\end{tabular}
\tablecomments{The parameters of our inferred GMM model, including the flattening (q) and rotation angles ($\alpha, \phi$) as well as the amplitudes for each GMM component. The amplitude for the inner Galaxy GMM ($A$; see Eqn. \ref{eqn:model}) and its 5 components ($A_{IG,j}$) along with the 6 halo amplitudes ($A_{H,j}$) are all given.  } 

\end{table}

We present the main result of our work, the Galactic distribution of RRL, in the form of the MLE of our 16-parameter GMM. The MLE and associated 1$\sigma$ uncertainties for the halo flattening and rotation are shown in Table \ref{tab:table1} as the parameters that are directly comparable to previous works \citep[e.g.,][]{Iorio2018,Han2022,Nibauer2025}. 

In order to visually asses the accuracy of our model, we compare the predicted sky projected density of RRL from our model to the observed sample. Figure \ref{fig:skymodel} shows three panels of sky projected density distribution with darker shading corresponding to higher densities. The top left panel is calculated by simply integrating our model along each line of sight. The top right panel shows the full forward model of the data by convolving the GMM with the empirical selection function. When compared to the sky projection of the observed data (bottom panel), we find generally qualitative agreement.  As the precision of the empirical selection function estimate depends on the sample size in each healpix and distance bin, it is worst close to the Galactic plane where dust extinction severely limits the RRL sample as shown in Figures \ref{fig:sky} and \ref{fig:sf}. Larger RRL catalogs are required to improve these results.

\subsection{Comparison to Literature Halo Models}

\begin{figure*}
    \centering
    \includegraphics[width=\linewidth]{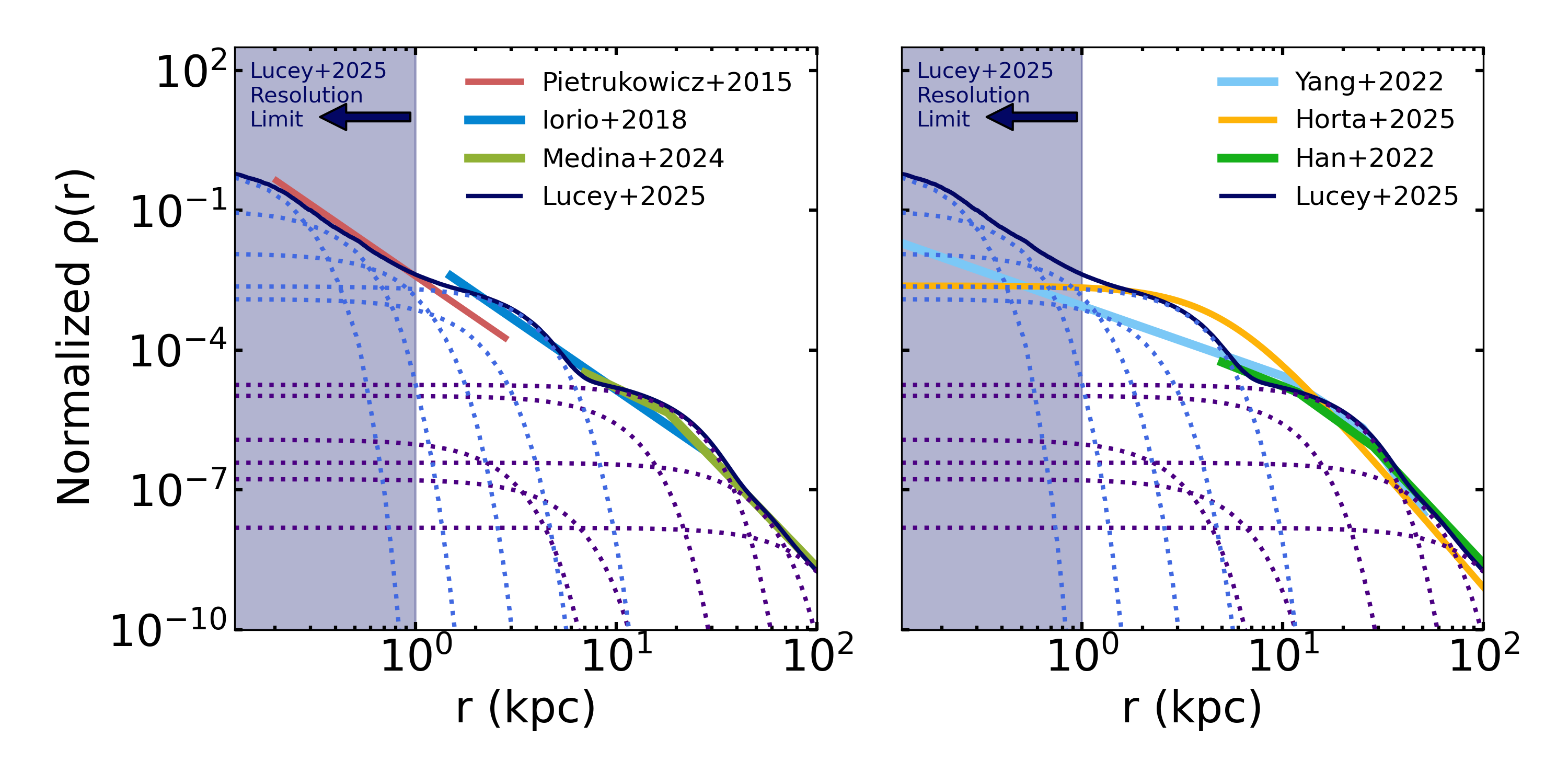}
    \caption{ The normalized halo density profile of RRL as a function of distance from the Galactic center ($r$) for our GMM compared to literature results using RRL (left) and other tracers (right).  In both panels, the light purple dotted lines indicate the 5 components of the inner Galaxy mixture model while the dark purple dotted lines show the halo components. The dark blue solid line is their sum.  The literature RRL profiles included in the left panel are \citet[][red]{Pietrukowicz2015}, \citet[][blue]{Iorio2018} and \citet[][green]{Medina2024}. The works in the right panel,   \citet[][yellow]{Horta2025}, \citet[][green]{Han2022}, and \citet[][light blue]{Yang2022},  use red giant stars.} 
    \label{fig:halo}
\end{figure*}

\begin{figure*}
    \centering
    \includegraphics[width=\linewidth]{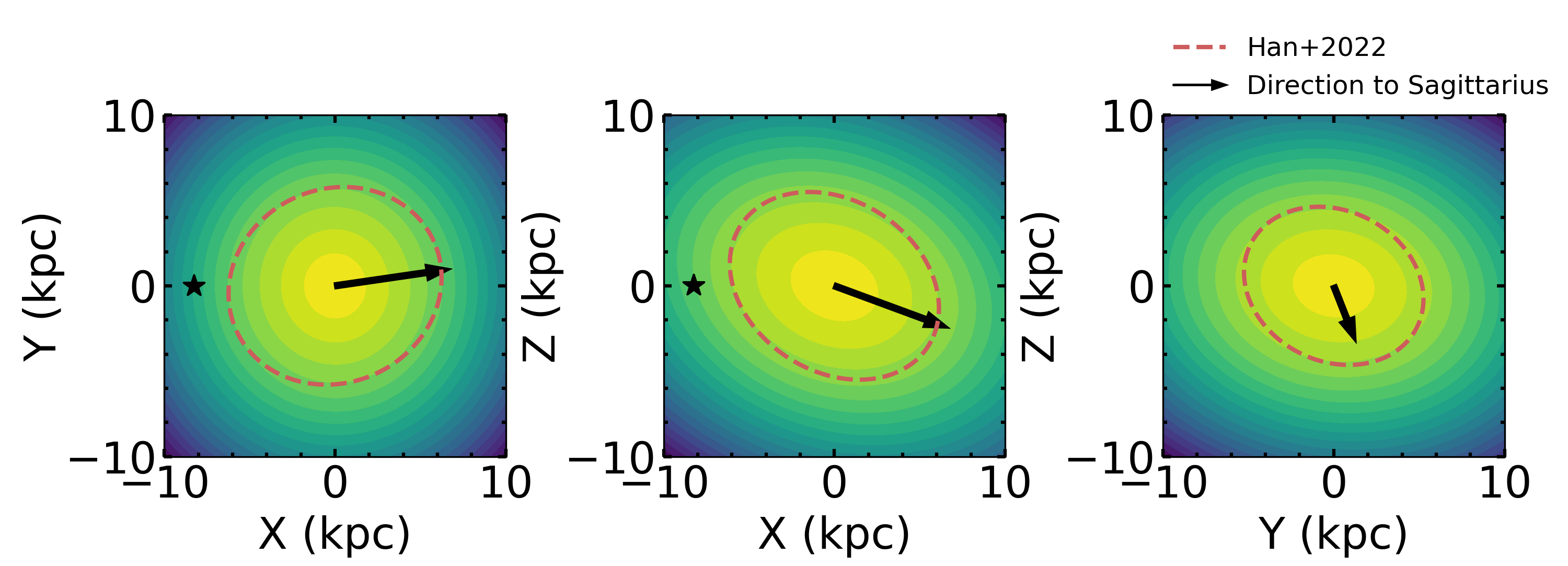}
    \caption{The inferred Galactic distribution of RRL as a GMM. The face-on distribution evaluated at Z=0~kpc is shown in the left panel. The middle and right panels show the edge-on distributions evaluated at X=0~kpc and Y=0~kpc, respectively. The location of the Sun is shown as a black star in the left and middle panels. In all panels, we include an ellipse (red-dashed line) which demonstrates the shape and tilt of the halo model from \citet{Han2022}. We also show the direction to Sagittarius which is located at (17.9, 2.6, $-$6.6)~kpc \citep{Vasiliev2021}. }
    \label{fig:contour}
\end{figure*}

Galactic halos are commonly described with power-law profiles, including cored, (doubly) broken, triaxial, and rotated adaptations \citep{Deason2011,Iorio2018,Han2022}. However, in this work, we use a GMM as an easily normalizable probability density and more flexible alternative to the power-law. This flexibility is especially crucial for modeling the inner Galaxy where the disk, halo and bulge all overlap. 

We show the normalized density as as function of Galactocentric radius (r) for the halo components of our GMM Galaxy in Figure \ref{fig:halo}, compared to results from the literature using RRL in the left panel and other tracers in the right panel. In each panel,  the 5 Gaussian halo components for the inner Galaxy GMM and the halo GMM are shown in light purple and dark purple dotted lines, respectively, with the sum shown in the solid dark blue line. In the left panel, the RRL profiles from the literature include results from \citet[red]{Pietrukowicz2015}, \citet[blue]{Iorio2018}, and \citet[green]{Medina2024}. These results are chosen to cover a wide range of radii, and includes both single and double power-law results, prioritizing recent work. The right panels includes the works of \citet[light blue]{Yang2022}, \citet[green]{Han2022} and \citet[][yellow]{Horta2025}. 

The literature results shown in the right panel are chosen to highlight the diversity of profiles measured using different stellar populations and methodologies to trace the stellar halo. For example, \citet{Yang2022} estimates the halo profile using direct orbit integration of K giant stars with galactocentric distances between 5 and 50 kpc. They remove disk stars using a selection in [Fe/H] and cylindrical velocity. Similar to our method, \citet{Han2022} use a Bayesian model to infer the halo profile using a sample of stars selected to be from the \textit{Gaia}-Enceladus-Sausage merger \citep{Belokurov2018,Helmi2018} using [$\alpha$/Fe] and [Fe/H] abundances as well as orbital eccentricity. \citet{Horta2025} uses a similar method to \citet{Han2022}, however they utilize a selection in [Mn/Mg], [Al/Fe], [Mg/Fe], and [Fe/H] to isolate the ``Heracles/proto-galaxy" structure that is generally confined to the inner Galaxy with apocenters<12~kpc. Our work improves upon these previous results by not performing a chemical or dynamical selection and instead directly modeling all of the available data including stars with Galactocentric radii as small as $\approx$0.2~kpc.

At Galactic radii $>$ 12 kpc, our results are generally consistent with previous results of a roughly $r^{-4}$ power law, measured both with RRL and other tracers \citep{Deason2011,Iorio2018,Han2022,Yang2022,Medina2024}. Within r$\approx$ 12 kpc, previous works have also found a flattening in the halo profile, with the power law decreasing to $r^{-1.5}$ \citep{Han2022,Yang2022,Medina2024}. Within $\approx$5 kpc of the Galactic center, we find that the inner Galaxy GMM begins to dominate the stellar density. We note that the amplitudes of the  halo GMM components at these radii are essentially zero, indicating that the dominant RRL population within 5 kpc of the Galactic center is not consistent with the rotated halo population. 

At Galactic radii >1~kpc and <5~kpc, we find a bump in the RRL density distribution. While the halo RRL slope has flattened between 5 and 12 kpc, the onset of the inner Galaxy RRL population causes a jump in the RRL distribution which again flattens approaching 1~kpc. Within 1~ kpc of the Galactic center our model agrees with results from \citet{Pietrukowicz2015}. However, our work is limited by selection function resolution at very small Galactic radii. Given our RRL sample size, the highest angular resolution we can achieve for estimating the empirical selection function is 7.32$^{\circ}$. This corresponds to $\approx$1 kpc at a the Galactic center assuming $R_{\odot}$=8.25~kpc. From Figure \ref{fig:skymodel}, it is also clear that our selection function is least accurate closest to the Galactic plane. However, our results robustly disagree with extrapolating this profile out to 5~kpc as suggested by \citet{PerezVillegas2017}.

In addition to comparing the slope of our halo model to previous works, we can also compare the rotation. Specifically, we compare our model to the tilted halo model of \citet{Han2022}, which was recently confirmed by an independent analysis in \citet{Nibauer2025}. In these works, they use a triaxial halo model, and relatedly, a different rotation matrix since their model does not have XY-symmetry, unlike ours. However, we can directly compare the pitch angles which both describe the tilt of the  z-axis towards the positive x-axis. We find a pitch angle of $18^{\circ}$,  in agreement with the halo mass distribution of \citet{Nibauer2025} and stellar density distribution of \citet{Han2022}. It is interesting to note that \citet{Han2022} specifically measure the stellar distribution of the \textit{Gaia}-Sausage-Enceladus system, which makes up the bulk of the local halo \citep{Belokurov2018,Helmi2018}. 

To visually compare our rotated halo with that of \citet{Han2022}, we plot density contours of the halo GMM in Figure \ref{fig:contour}, with the highest density in yellow and the lowest density in dark blue. The left panel shows the face-on slice evaluated at Z=0 kpc, while the middle and right panel show edge-on slices at X=0 kpc and Y=0 kpc, respectively. To guide the reader, we show the Sun's position (black star) in the left and middle panels at X=-8.25~kpc. The shape and tilt of our halo model agrees well with results from \citet{Han2022} shown as a red dashed line, especially in the middle panel which shows the X-Z slice. The left and right panels show slight disagreement with the \citet{Han2022} results, but this is likely because of the difference in rotation matrices, where our model preserves X-Y symmetry while \citet{Han2022} does not. 

In Figure \ref{fig:contour}, we also show an arrow that points to the location of the Sagittarius dwarf galaxy at (17.9, 2.6, $-$6.6)~kpc \citep{Vasiliev2021}. This aligns well with the rotated major-axis of the halo models. Sagittarius is a clear non-axisymmetric feature in our observed data (see Figure \ref{sec:Data}) so it is not unexpected that it could cause the tilt in our inferred halo model. \citet{Han2022} targets modeling the halo contribution of only the \textit{Gaia}-Enceladus-Sausage. Whether the tilt is caused by contamination from Sagittarius stars, or is a strange coincidence requires further investigation, including ongoing theoretical work testing the impact of assuming a single halo model when a galaxy is undergoing a Sagittarius-like merger versus explicitly modeling the second, less massive galaxy. \citet{Nibauer2025} models the dark matter halo tilt, however they assume an axisymmetric subtraction of stellar mass that does not include Sagittarius. This could cause the elongation of the ``missing" mass model towards the Sagittarius system.

For the inner Galaxy component, we measure an oblate shape with $q=1.31$. This is consistent with works measuring the shape of the dark matter halo within 20 kpc of the Galactic center using kinematic tracers \citep{Bowden2016,Bovy2016c}, including the work of \citet{Posti2019} which also found $q=1.3$. However, these works assume an oblate stellar halo or bulge for the non-disk component of the inner Galaxy. Further work to simultaneously study the stellar and dark matter mass distribution in the inner Galaxy is required to determine whether  the RRL are tracing the distribution of dark matter as predicted by cosmological zoom-in simulations \citep{Lucey2025}.

\section{Summary and Conclusions}\label{sec:Sum}
The stellar density distribution of the Milky Way provide clues to its formation and evolutionary history. Among the most challenging regions to characterize is the inner Galaxy where there are high levels of crowding and dust extinction. Yet, as the oldest component, disentangling its structure reveals crucial insight to the earliest epochs of galaxy evolution, including the role of dark matter \citep{El-Badry2018b,Rix2022,Lucey2025}.

To address these observational challenges, we use the standard candle nature of RRL stars to map the structure of the Milky Way with a hierarchical Bayesian GMM. This is the first time a GMM has been used to describe the stellar density distribution of the Milky Way, building on the success of this method in characterizing external galaxies including the power laws typically used to describe halos \citep{Cappellari2002,Hogg2013,Miller2021}. Crucially, we find the GMM provides the increased flexibility needed to model the inner Galaxy where the bulge, halo and disk all intersect. 

We use a total of 11 Gaussian components to model the Galactic distribution of RRL. All of the components are centered on the Galactic center. Five of the components are set to model the inner Galaxy with XY-covariances between 0.1-1.6 kpc and allowing the covariance in the z-direction to differ by a factor of $q$. We use a similar setup for the 6 components that model the halo, but with larger covariances (1.5-48 kpc) and allowing the z-axis to be rotated with respect to the Galactic disk. This framework yields three key findings about the Galactic distribution of RRL:

\begin{itemize}
    \item The inner Galaxy $(R\lesssim 10~\rm{kpc})$ is dominated by distinct prolate population with $q=1.31$, consistent with results for the shape of the dark matter distribution in the inner Galaxy \citep{Bowden2016,Posti2019}.

    \item The halo distribution is best described by a $r^{-4}$ power-law that flattens within 12 kpc of the Galactic center.
    
    \item We infer a flattening ($q$=0.70) and halo tilt in the RRL population ($\alpha=18^{\circ}$)  consistent with \citet{Han2022} and \citet{Nibauer2025}. We note for the first time that the rotated major axis points to the Sagittarius dwarf galaxy.

\end{itemize}

These results demonstrate that GMMs offer a powerful and flexible framework for modeling complex stellar distributions in the Milky Way. The main limit to our work is the sample size, especially close to the Galactic plane where we are severely limited by dust extinction. This especially impacts the selection function resolution and precision. Larger samples of RRL with well-calibrated distances, especially in high extinction regions will greatly improve this work, which the future Roman Galactic Plane Survey should provide \citep{Sanderson2024}.  Looking ahead, the next work in this series will apply this method to study how the halo structure varies as a function of stellar metallicity. As the relationship between the stellar halo's metallicity and structure is indicative of its origins \citep{Eggen1962,Searle1978}, such studies will provide insight into the formation history and chemical evolution of our Galaxy.

\appendix
\section{Comparing the data to the inferred model}
\begin{figure*}
    \centering
    \includegraphics[width=\linewidth]{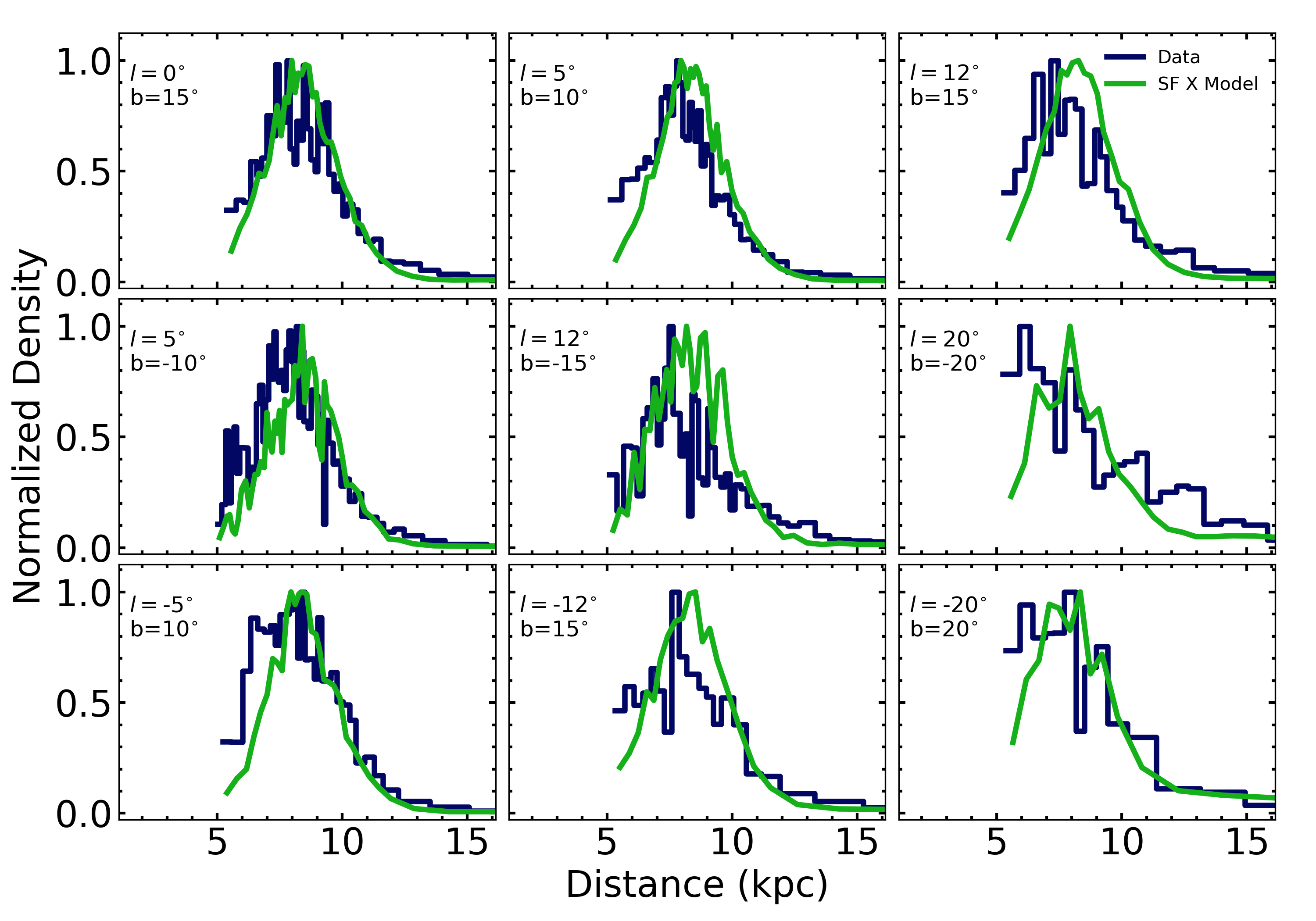}
    \caption{ The number density of the data (dark blue) compared to the selection function modified model (green) as a function of distance from the Sun along various lines of sight. The healpix and distance bins are set by the calculation of the selection function. Specifically, the distance bins are chosen so that for each line of sight each distance bin has the same number of stars used in the selection function calculation.    }
    \label{fig:modelfit}
\end{figure*}

To further asses the quality of our model fit to the observed data, we investigate the observed and predicted distribution of distances for a given line-of-sight in Figure \ref{fig:modelfit}. Given the complex selection function effects and distance uncertainties which are accounted for in our statistical model, it can be hard to directly compare the observed data distribution to our inferred model. This figure does not include the effect of our distance uncertainties, but we do show the predicted distribution of distances given our selection function and the model (green). For both the observed data (dark blue) and the model prediction, the binning matches that of the selection function calculation which is chosen so that for each healpix each distance bin has the same number of stars for the completeness calculation. The lines of sight chosen are all towards the Galactic center where the majority of our data lies. However, we do not show any lines of sight where $l<0^{\circ}$ and $b<0^{\circ}$, because the completeness is so low (see Figure \ref{fig:sf}) and there are two few stars.

In general, the selection function modified model fits the data well, especially given that the distance uncertainties. However, this figure also demonstrates how the accuracy and precision of our model is limited by the data and selection function calculation. While our model marginalizes over the distance uncertainties, we assume symmetric distance uncertainties and therefore in cases where we may be systematically underestimating the distances because of dust extinction towards the Galactic center, this will bias our results. In many of the panels, the peak in the observed data is at lower distances than the model and our data distribution contains more RRL at small distances than expected given the data and selection function. This is likely due to either the selection function underestimating the completeness at these distances or these distances being underestimated due to dust extinction. Larger RRL catalogs with improved distance estimates are required to improve these results.

\software{Astropy \citep{astropy:2013,astropy:2018},
Matplotlib \citep{matplotlib},
IPython \citep{ipython},
Numpy \citep{numpy}, 
Scipy \citep{scipy},
JAX  \citep{jax},
OTT-JAX \citep{ottjax},
Optax \citep{deepmindjax},
JAXopt \citep{jaxopt},
BlackJAX \citep{blackjax},
Corner \citep{corner}
luas \citep{Luas}
}    

\section*{Acknowledgements}

\begin{acknowledgments} This material is based upon work supported by the National Science Foundation under Award No. 2303831. 

This work used Bridges-2 at Pittsburgh Supercomputing Center through allocation PHY250006 from the Advanced Cyberinfrastructure Coordination Ecosystem: Services \& Support (ACCESS) program, which is supported by U.S. National Science Foundation grants \#2138259, \#2138286, \#2138307, \#2137603, and \#2138296. This work has made use of data from the European Space Agency (ESA) mission
{\it Gaia} (\url{https://www.cosmos.esa.int/gaia}), processed by the {\it Gaia}
Data Processing and Analysis Consortium (DPAC,
\url{https://www.cosmos.esa.int/web/gaia/dpac/consortium}). Funding for the DPAC
has been provided by national institutions, in particular the institutions
participating in the {\it Gaia} Multilateral Agreement.

\end{acknowledgments}

\bibliography{bibliography,bib_rrls}{}
\bibliographystyle{aasjournal}

\end{document}